# Electromechanical Oscillations in Bilayer Graphene

Muhammed Malik Benameur[1†], Fernando Gargiulo[2†], Sajedeh Manzeli[1†], Gabriel Autès[2], Mahmut Tosun[1], Oleg V. Yazyev[2], Andras Kis[1*]

[1]*Electrical Engineering Institute, École Polytechnique Fédérale de Lausanne (EPFL), CH-1015 Lausanne, Switzerland*
[2]*Institute of Theoretical Physics, École Polytechnique Fédérale de Lausanne (EPFL), CH-1015 Lausanne, Switzerland*

**Correspondence should be addressed to: Andras Kis, andras.kis@epfl.ch*
[†]*These authors contribute equally to this work*

**Nanoelectromechanical systems (NEMS) constitute a class of devices lying at the interface between fundamental research and technological applications. Integrating novel materials such as graphene into NEMS allows studying their mechanical and electromechanical characteristics at the nanoscale and addressing fundamental questions such as electron-phonon interaction and bandgap engineering. In this work, we integrate single and bilayer graphene into NEMS and probe the interplay between their mechanical and electrical properties. We show that the deflection of monolayer graphene nanoribbons results in a linear increase in their electrical resistance. Surprisingly, we observe oscillations in the electromechanical response of bilayer graphene. The proposed theoretical model suggests that these oscillations arise from quantum mechanical interference taking place due to the lateral displacement of graphene layers with respect to each other. Our work shows that bilayer graphene conceals unexpectedly rich and novel physics with promising potential in NEMS-based applications.**

Separating graphene[1] from the substrate by suspension permitted investigating its intrinsic electronic properties and allowed unveiling ultrahigh electron mobilities[2] due to reduced scattering and the observation of the fractional quantum Hall effect.[3,4] Similarly, suspension allowed incorporating graphene into NEMS in order to fabricate resonators[5,6] and opened the way to explore the interplay between the mechanical[7] and the electronic properties of graphene. However, during suspension, nanoscopic ripples form in order to ensure the thermodynamic stability of the two-dimensional crystal[8,9] and strain is inevitably introduced in suspended graphene. There is experimental evidence that strain strongly affects the physical properties of graphene.[10-14] As the crystalline symmetry is broken under strain it induces lifting of the two fold degeneracy of the optical phonon vibrational modes. This is observed as a splitting in the G peak in Raman spectra of strained graphene.[10,11] Very high pseudo-magnetic fields, reaching up to 300 T, have been predicted[12] and confirmed experimentally in highly strained graphene nanobubbles[13] and increased electrical resistance, reaching 5% under 3% tensile loads, has been reported[14] for graphene nanoribbons with widths ranging between 0.8 μm and 1.2 μm.

On the other hand, bilayer graphene has been reported to show appealing physical properties such as the possibility of opening and tuning an electronic bandgap[15] or engineering of quantum dots for single electron manipulation.[16] These properties make bilayer graphene emerge as a complementary material to single-layer graphene and open the way towards all-carbon based circuits where graphene could be used as high mobility conductor while bilayer graphene could in principle ensure electronic functionalities such as modulation or switching. The effect of strain on electrical transport in bilayer graphene has been investigated only theoretically.[17,18]

Here, we integrate mono- and bilayer graphene into NEMS in order to study the effect of strain on their transport properties. We report a study of the electrical response of graphene nanoribbons under local strain. Mono- and bilayer GNRs with widths between 60 nm and 300 nm have been investigated using nanoindentation techniques based on Atomic Force Microscopy (AFM) for high-resolution imaging and controlled deformation of the GNRs. Electrical conductance of the suspended GNR is measured simultaneously with mechanical deformation using a low-noise lock-in amplifier. Our samples consist of field-effect transistors based on suspended graphene nanoribbons (Figure 1a). The devices were fabricated using standard electron-beam lithography lift-off processes, followed by hydrofluoric acid (HF) wet etching and critical-point drying. Initial electrical characterization is performed with voltage sources connected in the configuration depicted in Figure 1b. We apply a bias $V_s$ to the source electrodes and a back-gate voltage $V_g$ to the degenerately doped silicon substrate. The drain current is measured using a current-voltage converter connected to a lock-in amplifier. The transfer and output characteristics of a monolayer and a bilayer suspended device is presented in Supplementary Section 1. All our devices have a linear $I_s - V_s$ characteristic typical of ohmic contacts. The gating dependence $I_s - V_g$ shows ambipolar dependence with the charge neutrality point $V_{CN}$ lying at the conductivity minimum and corresponding to the Dirac point.

Following the electrical characterization, we place the sample under AFM for imaging and nanoindentation. The imaging conditions are discussed in Supplementary Section 2. Prior to the nanoindentation experiment, the photodetector sensitivity and the AFM cantilever spring constant were calibrated (Supplementary Sections 3 and 4). Once the suspended GNR is located and the setup is calibrated, the AFM tip is positioned on top of the GNR for indentation. By moving the stage upwards against the cantilever and downwards far away from it, the GNR is deformed and then relaxed. This cycle of extension-retraction of the stage is represented in Figure 2a. During each deformation cycle, we simultaneously measure the current (upper graph) and the cantilever deflection, $D_{cantilever}$, (lower graph) as a function of the position of the stage, $Z_{piezo}$. We extract the deformation of the GNR ribbon at the point where load is applied by the AFM tip ($D_{GNR}$) from the expression $Z_{piezo} = D_{cantilever} + D_{GNR}$.[19] Detailed finite element modeling (FEM) has shown that the deflection of underetched contact areas (Figure 1b) can be neglected (Supplementary Section 7). During the experiment, an AC bias voltage with a root-mean-square (RMS) amplitude of 4 mV and a frequency of 8 kHz is applied to the GNR and the current flowing through it is monitored using a lock-in amplifier. We maintain the back-gate voltage connected to the ground, $V_g = 0$ V in order to exclude spurious effects due to changes in the capacitive coupling between the GNR and the gate as the GNR is deformed.

Figure 2a shows the electromechanical response of monolayer graphene. The electrical and mechanical responses are reproducible for both the extension and the retraction cycles. From the mechanical point of view, this proves that we are deforming the GNRs in the elastic regime, that no structural defects are introduced in the GNRs and that the GNRs are tightly anchored to the metallic pads (no slipping). Moreover, the electrical pads constitute a stable mechanical platform given their width (~ 2 μm) compared to the GNRs attached to them (~ 100 nm). From the electrical point of view, the reproducibility of the measurements proves that the interface between the GNRs and the metallic contacts does not deteriorate. The simultaneity of the measurements shows that the electrical response is tightly linked to the mechanical deformation. During the approach cycle and before mechanically contacting the GNRs the current is constant. The current undergoes variations only once the GNR is deformed ensuring that the observed current variations are of electromechanical origin.

In the case of monolayer graphene, we have performed electromechanical measurements on five devices with widths varying between 60 nm and 300 nm (Supplementary Section 5), resistances ranging between 10 kΩ and 100 kΩ with device mobilities ~1500 $cm^2$/Vs. We observe a linear decrease of the current under strain for both natural (device #1) and CVD graphene[20] (device #2 to device #5). We observe no opening of a bandgap in this deformation regime.

In order to compare the electromechanical response of the various GNRs, we represent the relative variation of the resistance $\Delta R/R_0$ as a function of nanoribbon deflection $D_{GNR}$ (Figure 2b). In almost all the devices (devices 1-4), we see an increase of electrical resistance as a result of strain ε, with a positive piezoresistive gauge factor $GF$ defined as $GF = (\Delta R/R_0)/\varepsilon$. This observed strain-induced resistance increase is in agreement with the previous reports on monolayer graphene[21] and is related to decreasing Fermi velocity and reduced mobility[14,21,22,23] Interestingly, one of our devices (device #5) shows a decrease in resistance as a function of strain, corresponding to a negative gauge factor. This rare behavior indicates that additional effects, possibly dependent on the lattice orientation, could modulate the electrical behavior of graphene under strain and warrants further theoretical modeling.

Because the sharp AFM tip with a radius of ~30 nm (Supplementary Section 4) introduces local strain in the nanoribbon center, resulting in non-uniform strain in the nanoribbon, we use finite element modeling to determine strain distribution in the ribbon, caused by the indenting AFM tip. Results show that, although in the vicinity of the AFM tip the dominant contribution to strain is from the local deformation under the sharp tip, in the rest of the ribbon, the strain is almost uniform and due to the overall vertical deflection of the ribbon (Supplementary Section 6). This allows us to estimate the upper limit on the gauge factor by taking into account the uniform strain induced by the vertical deflection while neglecting the contribution localized around the AFM tip. For devices 1-4, showing a positive gauge factor, we find an upper limit for GF of 8.8, in good agreement with previously published results[14,21] and at least twenty times lower than in semiconducting atomically thin layers of $MoS_2$[23] (a more detailed discussion is available in Supplementary Section 6). The same simulation results show that the highest achieved strains in our GNRs are ~5% (Supplementary Section 8).

We now turn to bilayer graphene devices. We have performed electromechanical measurements on nanoribbons fabricated from bilayer graphene with widths of 200 nm and 300 nm and resistances of ~ 50 kΩ and ~ 40 kΩ, respectively. The response of these two devices is represented in Figure 3 and in Supplementary Figure S11. Similarly to the samples of monolayer graphene, the mechanical response of bilayer devices is reversible, reflecting the elastic regime of deformation and the mechanical stability of the suspended bilayer GNRs. We can see from measurements of device current $I_d$ as a function of piezo scanner extension, Figure 3a, that the electromechanical response of bilayer GNRs shows two main features. Similarly to monolayer GNRs, we see an overall increase of resistance as a function of nanoribbon deflection $D_{GNR}$, Figure 3b. In addition, we observe pronounced oscillations superposed on the background of increasing resistance. These oscillations in the electrical response exhibit the same qualitative behavior in the extension and retraction cycles (Figure 3a). However, the oscillations from both cycles are slightly out of phase while the amplitudes have the same order of magnitude. The current before and after deformation remains unchanged. This confirms that the device has not deteriorated during deformation and proves the electromechanical origin of the observed oscillations. In order to compare our measurements, we consider only the response from the extension cycles. By performing successive deformation, we see that the oscillations in the relative change in resistance are highly repeatable and qualitatively similar for both devices (Figure 3b and Supplementary

Figure S11), with a peak to peak amplitude of ~4% and a frequency that increases as the nanoribbon is deflected.

The overall, background increase of resistance in bilayer devices can be explained in terms of the piezoresistive response, just as in the case of monolayer devices. The gauge factor of bilayer devices is, however, lower than monolayer devices. Calculations by Wong et al.[24] show that tensile strain can increase interlayer interactions in bilayer graphene, which could partially compensate decreasing intralayer interactions, resulting in a smaller gauge factor for bilayer graphene.

Clearly, interlayer interactions are at the origin of the striking observation of electromechanical oscillations. We propose a simple theoretical model capable of quantitatively reproducing the observed behavior. The model assumes that the AFM tip action causes finite lateral displacement (sliding) of the individual graphene layers with respect to each other. This lateral displacement is expected due to the weak van der Waals force between the two graphene layers.[25,26] There is extensive evidence in the literature that lateral displacement of graphene layers with respect to each other takes place in various types of scanning probe microscopies.[27-36] The relative displacement of a single graphene layer on another graphene layer is even more likely when accompanied by deformation of the layers.[37] In our experiment, the AFM tip deforms the bilayer GNRs which leads to an increase in the elastic energy of the system. Therefore, the necessity for lowering the energy of the system results in the relative displacement of the layers. The lateral displacement, however, disturbs the AB-stacking of the two layers which is the energetically preferred configuration of bilayer graphene.[38] The system then reduces the interlayer binding energy through formation of a "domain wall"-like transition region separating two AB-stacked domains, as described by the Frenkel-Kontorova model[39] (see Figure 4a). The transition region essentially accommodates the transition to the incommensurate phase and in turn allows the AB-stacking in the two regions of GNR on each side of the "domain wall" that would appear as a localized bulge, or wrinkle, for the reasons discussed below.[38,39] The width of these transition regions, typically a few nanometers according to experimental observations[29] and our numerical estimates (see Supplementary Section 11), is defined by the balance between the total strain energy and the interlayer binding energy.[39,40] Such boundaries occurring between AB- and BA-stacked regions have recently been observed in the misoriented multilayer graphenes by numerous groups.[26,29,33,35,36,38] Displacement of stacking domain boundaries and manipulations and creation of wrinkles in scanning tunneling microscopy (STM) was demonstrated.[27,32,34,36] Displacement and removal of wrinkles with an AFM tip was also shown experimentally,[28,30,31] and the details of this process were investigated by means of realistic simulations.[41]

We note, however, that the effective width of the transition region within the two individual graphene layers is different. Moreover, this effective width difference $\Delta W$ will vary as the two graphene layers slide against each other due to the action of the AFM tip. If no transition region was initially present in the sample, $\Delta W$ corresponds to the lateral displacement of one graphene layer with respect to another upon indentation. In the transition region, the layers are electronically decoupled either due to their incommensurate stacking[42] for small values of $\Delta W$, or due to enlarged interlayer distance for larger $\Delta W$ giving rise to a wrinkle (localized bulge) in one of the layers of bilayer graphene, as illustrated in Figure 4a. We suggest that the observed electromechanical oscillations can be explained from the point of view of quantum interference phenomena due to path difference $\Delta W$ of the charge carriers in the decoupled graphene layers. Increasing the strain leads to higher amount of local corrugation and changes $\Delta W$ which would cause constructive or destructive interference.

In order to verify this hypothesis, we perform numerical simulations of electronic transport in a model bilayer graphene device. The simulations are performed in the ballistic regime because of the few nanometers width of the transition region. The methodology is based on a tight-binding Hamiltonian and non-equilibrium Green's function formalism (see Supplementary Section 10 for details). Without loss of generality, the individual graphene layers are assumed to be fully decoupled (zero interlayer hopping integrals) in the transition region. Our model device is a bilayer graphene with the zigzag direction aligned to the transport direction and is periodic in the perpendicular (armchair) direction. We investigated armchair direction of domain boundaries as this orientation was found to be dominant for the case of closely related AB-BA stacking domain boundaries extensively investigated using transmission electron microscopy.[38] We assume that the sample is large enough that edge effects do not affect significantly transport properties. Furthermore, considering the in plane isotropic elasticity of graphene, the crystallographic orientation of the transport channel does not affect the electromechanical behavior. Figure 4b shows the calculated charge-carrier transmission probabilities as a function of energy $E$ and momentum parallel to the transition region, $k_{\|}$, for various carrier path differences $\Delta W = na$ ($n \in \mathbb{Z}$), $a$ = 0.246 nm – the lattice constant of graphene). The trivial case of $\Delta W = 0$ (no transition region, i.e. pristine bilayer graphene) reveals the massive character of Dirac fermions in bilayer graphene. Finite path differences $\Delta W$ result in significant amount of backscattering developing a clear sub-band sequence resulting from the quantum confinement of massless Dirac fermions in the transition region (indicated by the dashed line in the last panel of Figure 4b). Most importantly, configurations characterized by $\Delta W = 3ma$ ($m \in \mathbb{Z}$) show enhanced transmission due to constructive interference as the wavenumber of Dirac fermions in graphene $k = \frac{4\pi}{3a}$. Therefore, One period of oscillations corresponds to the deformation-induced lateral displacement of $3a$ = 0.74 nm.

In order to gain further insight, we compare the calculated resistance with actual experimental observation. A quantitative comparison requires accounting for the role of contacts as well as for the diffusive transport in the rest of device. Both factors, below collected in a single value $R_c$, act as a 'bottleneck' in a realistic device and are thus responsible for most of its total resistance. We obtained $R_c = 41k\Omega$ by fitting both the average value of the calculated resistance and the magnitude of oscillations to the experimental data reported in Figure 3b for device #7. In Fig. 4c oscillations in resistance can be seen clearly with a constant period proportional to $3a$. The direct comparison of Figure 3b and Figure 4c reveals a quantitative agreement with experiments, except for the fact that in simulations $\Delta W = 0$ corresponds to the minimum of resistance (no backscattering) while in experiments resistance oscillates reaching both higher and lower values compared to the zero-displacement point. This implies that the stacking domain boundary was already present before indentation. In other words, the origin in experimental resistance curves corresponds to a finite value of $\Delta W$ in Figure 4c.

In summary, we have investigated the electromechanical response of mono- and bilayer graphene nanoribbons (GNRs). Monolayer graphene devices show an increase in resistance under strain which is related to a change in the Fermi velocity under strain. Within our experimental conditions at room temperature, we observe neither a spectral nor a transport band gap larger than 4 meV for uniaxial strain under 5%, in agreement with theoretical predictions by Pereira et al.[43] Additionally, we report on the electromechanical response of bilayer graphene, which shows a superposition of linear response with oscillations in the resistance. The observed oscillations are reproduced within the framework of a simple theoretical model and we show that they can be explained as an interference phenomenon taking place between the two graphene layers. It is interesting to note that this interference effect is observed at room temperature, which is quite rare in the wider context of electronic interference phenomena. The successful integration of bilayer graphene into NEMS devices shows that bilayer graphene conceals unexpectedly rich physics and that bilayer-based NEMS could be a new interesting system for studying symmetry breaking in graphene and for studying electronic interference phenomena at room temperature.

## MATERIALS AND METHODS

Single and bilayer sheets of graphene have been exfoliated from commercially available crystals of graphite (NGS Naturgraphite GmbH) using the scotch-tape micromechanical cleavage technique.[1] CVD graphene sheets have been obtained by growth of graphene on a 99.8% pure, annealed, copper foil (Alfa Aesar) following a two-step growth recipe.[44] The graphene is then transferred on top of Si substrate covered with a 270 nm thick $SiO_2$ layer.[20] The surface of the samples is imaged using an optical microscope (Olympus BX51 M) equipped with a color camera (AVT Pike F-505C). Electrical characterization of the devices is carried out using Agilent E5270B parameter analyzer and a home-built shielded probe station with micromanipulated probes. AFM imaging and electromechanical nanoindentation experiments are performed using a home-built set-up combining the Asylum Research Cypher AFM with the low noise lock-in amplifier (SRS-830). We use nonconductive Si AFM tips, model NSC36/AlBS from MikroMash (Supplementary Section 4).

## ACKNOWLEDGEMENTS

Device fabrication was carried out in the EPFL Center for micro-nanotechnology (CMI). We thank K. Lister and Z. Benes for their support with e-beam lithography. M.B. and A.K. acknowledge financial support from the Swiss National Science Foundation (Grant no. 200021_122044). This work was financially supported by funding from the European Union's Seventh Framework Programme FP7/2007-2013 under Grant Agreement No. 318804 (SNM) and was carried out in frames of the Marie Curie ITN network "MoWSeS" (grant no. 317451). F.G., G.A. and O.V.Y. acknowledge support from the Swiss National Science Foundation (grant no. PP00P2_133552). We acknowledge funding by the EC under the Graphene Flagship (grant agreement no. 604391). Computations have been performed at the Swiss National Supercomputing Centre (CSCS) under project s515.

### Supporting Information

The Supporting Information document contains data on electrical characterization of devices, images of suspended graphene nanoribbons, details on AFM probe and detector calibration, details of devices geometries and preparation method, results on FEM modeling of strain distribution, the gauge factor and boundary conditions, discussion of the mechanical failure and reproducibly of electromechanical oscillations, description of the simulation methodology and estimation of the width of transition regions. This material is available free of charge via the Internet at http://pubs.acs.org.

### Author Contributions

MB fabricated the devices, performed the experiment and analyzed the data. SM fabricated devices, performed the experiment on an additional set of devices and carried out FEM simulations. AK designed the experiment. FG, GA and OY performed theoretical calculations. MT performed CVD graphene growth. MB, SM, FG, GA, OY and AK wrote the manuscript.

## References

1 Novoselov, K. S. *et al.* Electric field effect in atomically thin carbon films. *Science* **306**, 666-669, doi:10.1126/science.1102896 (2004).
2 Bolotin, K. I. *et al.* Ultrahigh electron mobility in suspended graphene. *Solid State Communications* **146**, 351-355 (2008).
3 Bolotin, K. I., Ghahari, F., Shulman, M. D., Stormer, H. L. & Kim, P. Observation of the fractional quantum Hall effect in graphene. *Nature* **462**, 196-199, doi:http://www.nature.com/nature/journal/v462/n7270/suppinfo/nature08582_S1.html (2009).
4 Du, X., Skachko, I., Duerr, F., Luican, A. & Andrei, E. Y. Fractional quantum Hall effect and insulating phase of Dirac electrons in graphene. *Nature* **462**, 192-195 (2009).
5 Bunch, J. S. *et al.* Electromechanical Resonators from Graphene Sheets. *Science* **315**, 490-493, doi:10.1126/science.1136836 (2007).
6 Chen, C. *et al.* Performance of monolayer graphene nanomechanical resonators with electrical readout. *Nature Nanotechnology* **4**, 861-867, doi:http://www.nature.com/nnano/journal/v4/n12/suppinfo/nnano.2009.267_S1.html (2009).
7 Lee, C. *et al.* Frictional Characteristics of Atomically Thin Sheets. *Science* **328**, 76-80, doi:10.1126/science.1184167 (2010).
8 Meyer, J. C. *et al.* The structure of suspended graphene sheets. *Nature* **446**, 60-63 (2007).
9 Bao, W. Controlled ripple texturing of suspended graphene and ultrathin graphite membranes. *Nature Nanotech.* **4**, 562-566 (2009).
10 Huang, M. *et al.* Phonon softening and crystallographic orientation of strained graphene studied by Raman spectroscopy. *Proceedings of the National Academy of Sciences* **106**, 7304-7308, doi:10.1073/pnas.0811754106 (2009).
11 Mohiuddin, T. M. G. *et al.* Uniaxial strain in graphene by Raman spectroscopy: G peak splitting, Grüneisen parameters, and sample orientation. *Physical Review B* **79**, 205433 (2009).
12 Guinea, F., Katsnelson, M. I. & Geim, A. K. Energy gaps and a zero-field quantum Hall effect in graphene by strain engineering. *Nat Phys* **6**, 30-33 (2010).
13 Levy, N. *et al.* Strain-Induced Pseudo-Magnetic Fields Greater Than 300 Tesla in Graphene Nanobubbles. *Science* **329**, 544-547, doi:10.1126/science.1191700 (2010).
14 Huang, M., Pascal, T. A., Kim, H., Goddard, W. A. & Greer, J. R. Electronic−Mechanical Coupling in Graphene from in situ Nanoindentation Experiments and Multiscale Atomistic Simulations. *Nano Letters* **11**, 1241-1246, doi:10.1021/nl104227t (2011).
15 Zhang, Y. *et al.* Direct observation of a widely tunable bandgap in bilayer graphene. *Nature* **459**, 820-823 (2009).
16 Allen, M. T., Martin, J. & Yacoby, A. Gate-defined quantum confinement in suspended bilayer graphene. *Nat Commun* **3**, 934, doi:10.1038/ncomms1945 (2012).
17 Mucha-Kruczyński, M., Aleiner, I. L. & Fal'ko, V. I. Strained bilayer graphene: Band structure topology and Landau level spectrum. *Physical Review B* **84**, 041404 (2011).
18 Mucha-Kruczyński, M., Aleiner, I. L. & Fal'ko, V. I. Landau levels in deformed bilayer graphene at low magnetic fields. *Solid State Communications* **151**, 1088-1093, doi:10.1016/j.ssc.2011.05.019 (2011).
19 Tombler, W. *et al.* Reversible electromechanical characteristics of carbon nanotubes underlocal-probe manipulation. *Nature* **405**, 769-772, doi:doi:10.1038/35015519 (2000).

20 Li, X. *et al.* Large-Area Synthesis of High-Quality and Uniform Graphene Films on Copper Foils. *Science* **324**, 1312-1314, doi:10.1126/science.1171245 (2009).
21 Smith, A. D. *et al.* Electromechanical Piezoresistive Sensing in Suspended Graphene Membranes. *Nano Letters* **13**, 3237-3242, doi:10.1021/nl401352k (2013).
22 Choi S.-M., Jhi S.-H & Son Y.-W. Effects of strain on electronic properties of graphene. *Physical Review B* **81** (2010).
23 Manzeli, S., Allain, A., Ghadimi, A. & Kis, A. Piezoresistivity and Strain-induced Band Gap Tuning in Atomically Thin MoS2. *Nano letters*, doi:10.1021/acs.nanolett.5b01689 (2015).
24 Wong, J.-H., Lin, F., Bi-Ru, W. & Ming, F. Strain Effect on the Electronic Properties of Single Layer and Bilayer Graphene. *The Journal of Physical Chemistry C* **116**, 8271–8277, doi:10.1021/jp300840k (2012).
25 Zheng, Q. e. a. Self-Retracting Motion of Graphite Microflakes. *Phys. Rev. Lett. 100, 067205 (2008).* (2015).
26 Brown, L. *et al.* Twinning and Twisting of Tri- and Bilayer Graphene. *Nano Lett.* **12**, 1609–1615 doi:10.1021/nl204547v (2012).
27 Xu P. & Yurong Yang, D. Q., S. D. Barber, J. K. Schoelz, M. L. Ackerman, L. Bellaiche, and P. M. Thibado. Electronic transition from graphite to graphene via controlled movement of the top layer with scanning tunneling microscopy. *Phys. Rev. B* **86**, doi:, 085428 – Published (2012).
28 Camara N & Jean-Roch Huntzinger, G. R., Antoine Tiberj, Narcis Mestres, Francesc Pérez-Murano, Philippe Godignon, and Jean Camassel. Anisotropic growth of long isolated graphene ribbons on the C face of graphite-capped $6H$-SiC. *Phys. Rev. B* **80**, doi:125410 (2009).
29 Alden, J. S. *et al.* Strain solitons and topological defects in bilayer graphene. *Proceedings of the National Academy of Sciences of the United States of America* **110**, 11256-11260, doi:10.1073/pnas.1309394110 (2013).
30 Li, Z. *et al.* Spontaneous Formation of Nanostructures in Graphene. *Nano letters* **9**, 3599-3602, doi:10.1021/nl901815u (2009).
31 Lin, Q.-Y. *et al.* Stretch-Induced Stiffness Enhancement of Graphene Grown by Chemical Vapor Deposition. *ACS nano* **7**, 1171-1177, doi:10.1021/nn3053999 (2013).
32 Lalmi, B. *et al.* Flower-Shaped Domains and Wrinkles in Trilayer Epitaxial Graphene on Silicon Carbide. *Scientific reports* **4**, doi:doi:10.1038/srep04066 (2014).
33 Lin, J. *et al.* AC/AB Stacking Boundaries in Bilayer Graphene. *Nanolett.* **13(7)**, 3262-3268, doi:10.1021/nl4013979 (2013).
34 Xu, K., Heath, J. R., Peigen, C. & James, R. Scanning Tunneling Microscopy Characterization of the Electrical Properties of Wrinkles in Exfoliated Graphene Monolayers. *Nano Lett.* **9**, 4446-4451, doi:10.1021/nl902729p (2009).
35 Butz, B. *et al.* Dislocations in bilayer graphene. *Nature* **505**, 533-537, doi:doi:10.1038/nature12780 (2013).
36 Yankowitz, M. *et al.* Electric field control of soliton motion and stacking in trilayer graphene. *Nature materials* **13**, 786-789, doi:doi:10.1038/nmat3965 (2014).
37 Brown, L. *et al.* Twinning and Twisting of Tri- and Bilayer Graphene. *Nano Letters* **12**, 1609-1615, doi:10.1021/nl204547v (2012).
38 Gong, L. *et al.* Reversible Loss of Bernal Stacking during the Deformation of Few-Layer Graphene in Nanocomposites. *ACS nano* **7**, 7287-7294, doi:10.1021/nn402830f (2013).
39 Popov, A. M., Lebedeva, I. V., Knizhnik, A. A., Lozovik, Y. E. & Potapkin, B. V. Commensurate-incommensurate phase transition in bilayer graphene. *Physical Review B* **84**, 045404 (2011).

40 Cambridge, D. H. E. G. s. P. U. o. C. & D, J. *Introduction to Dislocations, Fifth Edition*. (Butterworth-Heinemann, 2011).
41 Ye, Z., Martini, Z. Y., Chun, T., Yalin, D. & Ashlie. Role of wrinkle height in friction variation with number of graphene layers. *Journal of Applied Physics* **112**, doi:doi:10.1063/1.4768909 (2012).
42 Lopes dos Santos, J. M. B., Peres, N. M. R. & Castro Neto, A. H. Graphene Bilayer with a Twist: Electronic Structure. *Physical Review Letters* **99**, 256802 (2007).
43 Pereira, V. M., Castro Neto, A. H. & Peres, N. M. R. Tight-binding approach to uniaxial strain in graphene. *Physical Review B* **80**, 045401 (2009).
44 Li, X. *et al.* Graphene Films with Large Domain Size by a Two-Step Chemical Vapor Deposition Process. *Nano Letters* **10**, 4328-4334, doi:10.1021/nl101629g (2010).

## Figures

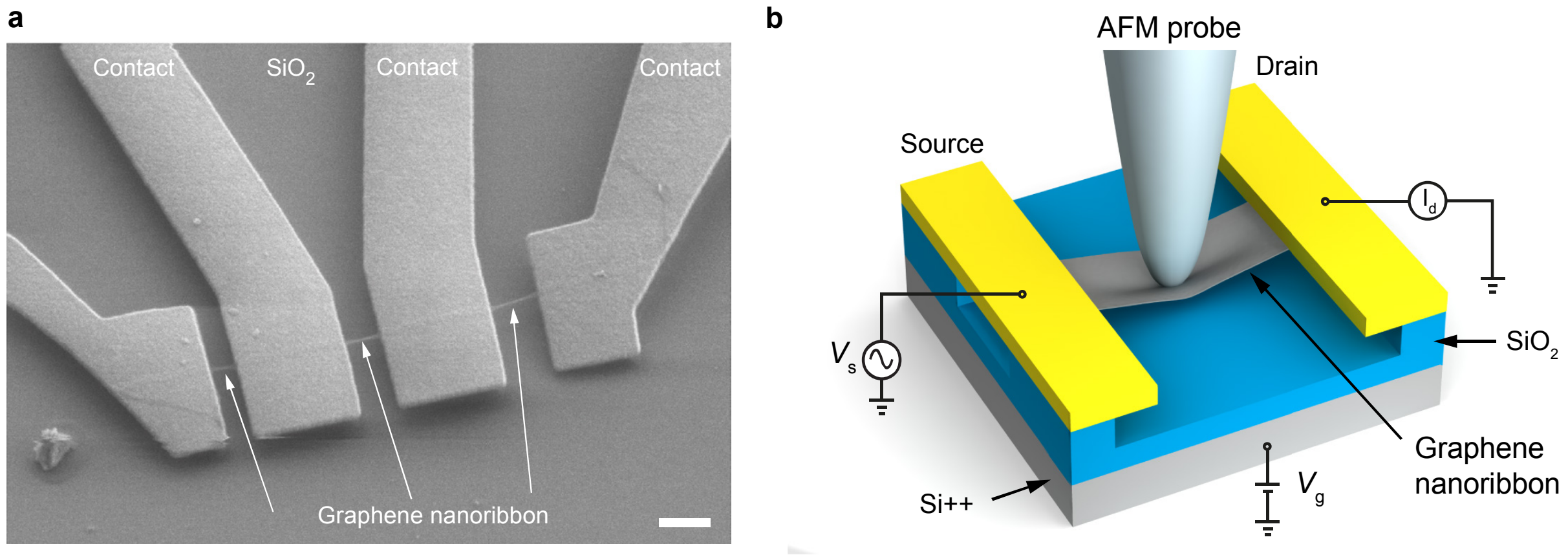

Figure 1. Device and experimental setup. **a**, Scanning electron microscope (SEM) image of the device. A 60 nm wide graphene nanoribbon is suspended above a substrate and contacted by electrodes. Scale bar is 500 nm long. **b**, Schematic drawing of the experimental setup and geometry. The suspended graphene ribbon is deformed in the center using an AFM probe attached to a piezo scanner. The vertical displacement of the scanner $Z_{piezo}$ results in the deflection of the cantilever $D_{cantilever}$ and nanoribbon deflection $D_{GNR}$. The device is biased by an AC voltage with an RMS amplitude of 4mV. The resulting drain current $I_d$ is monitored using a lock-in amplifier.

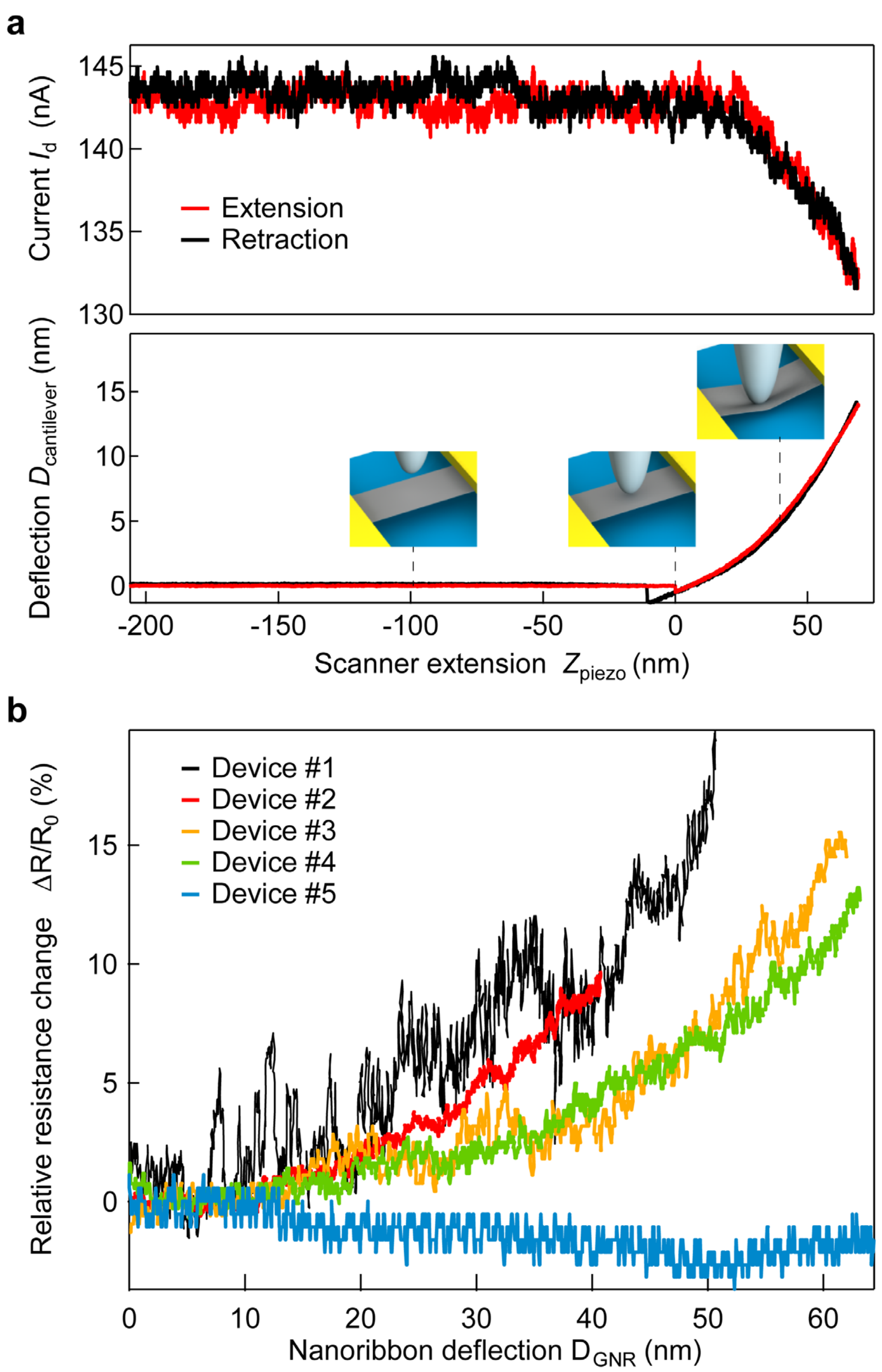

Figure 2. Electromechanical response of monolayer graphene. **a,** Electromechanical experiment shows simultaneous measurements of the current (upper curve) and the cantilever's deflection (lower part) as a function of the piezo scanner extension. The electromechanical response is reproducible for both extension (red) and retraction (black) curves. The measurement is performed for an AC voltage with an RMS amplitude of 4 mV and with the grounded back-gate. Further analysis (see equations in the main text) allows extraction of **b,** relative variation of the resistance as a function of the nanoribbon deflection. All monolayer graphene devices show a response with varying slopes depending on the GNR width. In most cases, the resistance increases under strain, however, we observed one case of decreasing resistance under strain (blue curve, device #5).

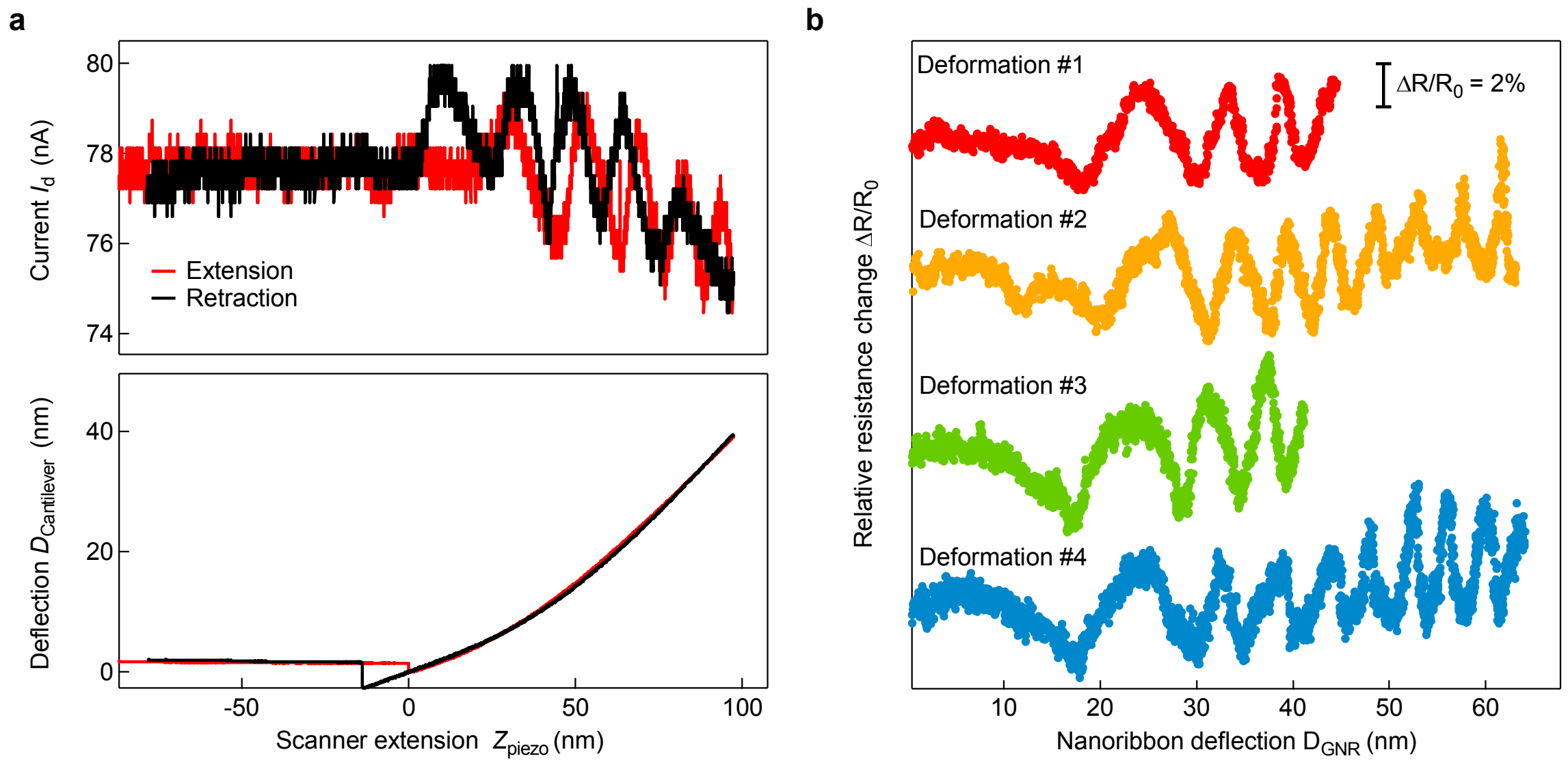

Figure 3. Electromechanical response of bilayer graphene. **a,** Simultaneous measurements of the current (upper curve) and the cantilever's deflection (lower part) as a function of the piezo-scanner extension show oscillations in the electrical response of bilayer GNRs. Oscillations are reproducible and slightly out-of-phase for both extension and retraction cycles. The measurement is performed for an AC voltage with an RMS amplitude of 4 mV and with the back-gate grounded. **b,** Relative resistance of a bilayer graphene nanoribbon as a function of deflection for several successive cycles of mechanical deformation. Curves are offset for clarity. Oscillations in resistance with an amplitude of ~2% are superposed on a slowly increasing background.

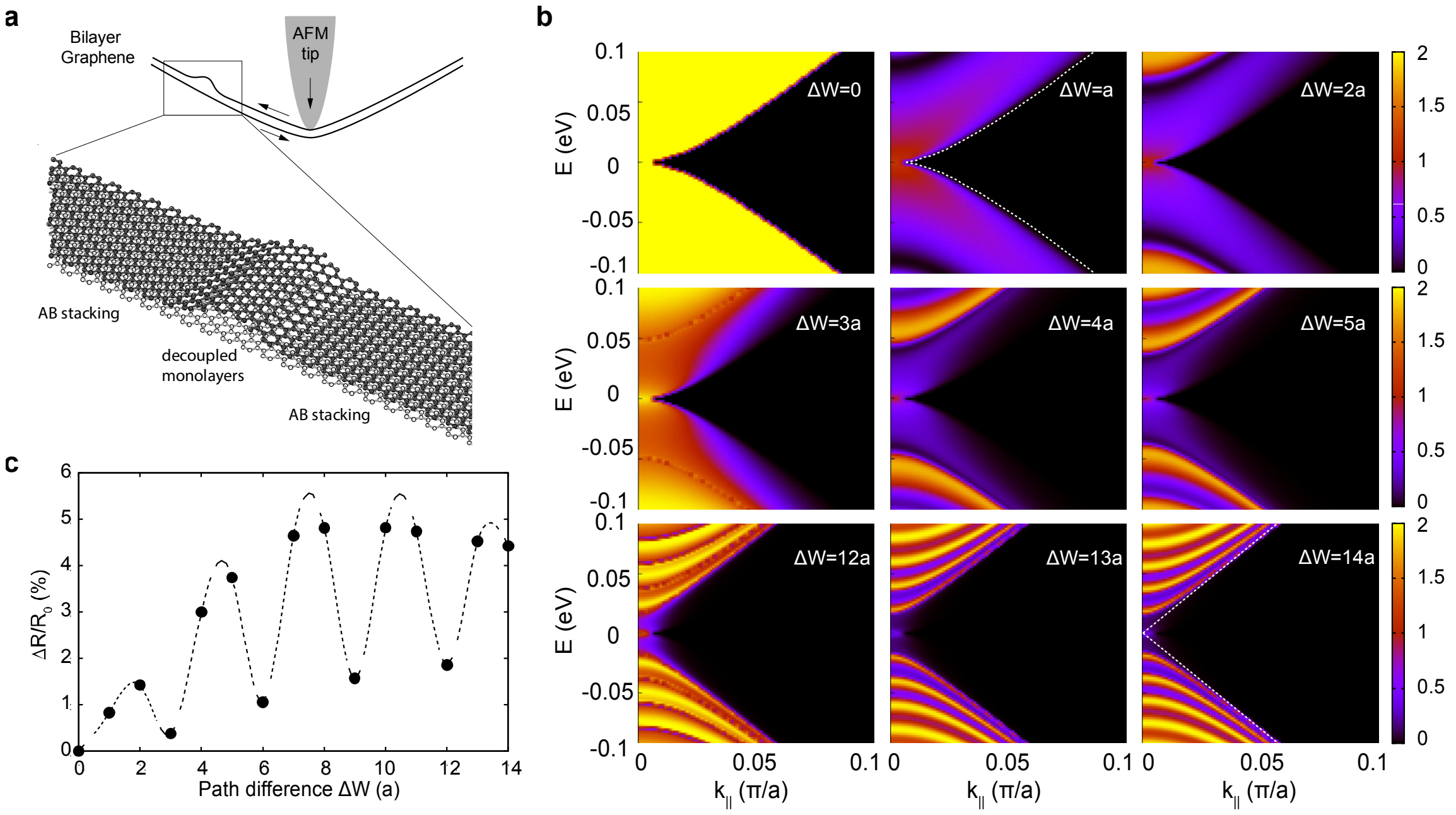

Figure 4. Theoretical simulations of charge-carrier transport. **a,** Schematic illustration of the lateral shift of individual graphene layers with respect to each other subjected to the AFM tip action. The AB-stacked bilayer graphene domains are separated by a region of decoupled monolayers of different effective width. **b,** Calculated charge-carrier transmission probability across a region of decoupled graphene monolayers as a function of $E$ and $k_{||}$ for various charge-carrier path differences $\Delta W$, given in units of lattice constant of graphene $a$. The dashed lines show the contours of the massive Dirac fermion band of bilayer graphene and the massless Dirac cone of monolayer graphene, respectively. **c,** Relative electrical resistance $\Delta R/R_0$ of the simulated nanoelectromechanical device based on a bilayer GNR with a width of 50 nm under $V_{\text{bias}}$ = 4 mV with a contact resistance $R_c$ = 41 kOhm as a function of charge carrier path difference $\Delta W$ given in units of lattice constant of graphene *a.* The line is a guide to the eye.